\documentclass[a4paper,11pt]{article}
\pdfoutput=1 

\usepackage{jcappub} 

\usepackage[T1]{fontenc}
\newcommand{\Lagr}{\mathcal{L}}
\newcommand{\kessence}{\mathcal{K}}
\newcommand{\Mpl}{M_{Pl}}
\newcommand{\phdt}{\dot{\phi}}
\newcommand{\hds}{H_{ds}}
\newcommand{\gev}{\, \text{GeV}}
\newcommand{\smalld}{\text{d}}
\newcommand{\laplace}{\mathcal{F}}
\newcommand{\ghostfree}{\mathcal{G}}
\newcommand{\sss}{c_s^2}

\title{A minimal self-tuning model to solve the cosmological constant problem}

\author{Arnaz Khan}
\author{and Andy Taylor}
\affiliation{Institute for Astronomy, University of Edinburgh, \\Royal Observatory, Blackford Hill, Edinburgh, EH9 3HJ, U.K.}

\emailAdd{arnaz.khan@ed.ac.uk}
\emailAdd{ant@roe.ac.uk}

\abstract{The expansion of the Universe is observed to be accelerating, with the simplest solution being a classical cosmological constant. However, this receives contributions from the quantum vacuum, which are predicted to be many orders of magnitude larger than observations, and suffers from radiative instabilities requiring repeated fine-tuning. In this paper we present a minimal, self-tuning scalar field model that can dynamically cancel a large quantum vacuum energy, avoiding Weinberg’s No-Go Theorem, and produce accelerated de Sitter expansion at a lower energy scale as a solution to the problem. Our minimal model, which contains a non-canonical kinetic energy and a linear potential, belongs to the Kinetic Gravity Braiding sub-class of Horndeski theory which is not observationally excluded, and lies outside of the known Fab-Four or Well-Tempered models. We find analytic solutions in the limits of slow-roll and fast-roll, and numerically solve the equations of motion to illustrate our model. We show that the model allows for a matter dominated era, and that the attractor solution is stable under a phase transition in the vacuum energy density. We also consider the energy-scales required to match observations. Our model shows the existence of a wider class of successful self-tuning models than previously assumed.}

\begin{document}
\maketitle
\flushbottom

\section{Introduction}
\label{sec:intro}

The discovery of the accelerated expansion of the Universe \cite{riess1998observational}, \cite{perlmutter1999measurements} presents a serious challenge to cosmology and physics. It implies an effective dominant, negative-pressure component to the Universe, or an extension of gravity to allow an increasing field strength, to destabilise the expansion. A natural candidate is the classical cosmological constant with its amplitude matched to observations, but this receives contributions from quantum fluctuations in the vacuum. The amplitude of these fluctuations can be estimated but is many orders of magnitude larger than observations imply. Moreover, the cosmological constant has radiative instabilities, meaning any correction to its value through renormalisation to match observations requires fine-tuning at every loop-order due to its sensitivity to unknown UV physics \cite{padilla2015lectures}, \cite{martin2012everything}. In addition, there is a contribution from classical fields whose potential can vary significantly between phase transitions. This implies that Nature has a high-energy solution to all of this, and we only see a low-energy residual effect. This is known as the Cosmological Constant Problem.

One effective solution to the Cosmological Constant Problem could be for a single scalar field coupled to the metric to eat up the large vacuum energy density and produce an effective cosmological constant at the required energy scale $(\sim\text{meV}^4)$. However, Weinberg’s No-Go Theorem \cite{weinberg1989cosmological} (see also \cite{padilla2015lectures} for an elegant discussion) shows that under reasonable conditions any scalar field model that relaxes to a constant vacuum expectation value and cancels the vacuum energy will need to be fine-tuned.

One can avoid the No-Go Theorem by relaxing its assumptions, in particular, by ensuring that the scalar field remains dynamic throughout its evolution. This was originally proposed for a self-tuning model \cite{charmousis2012self} which evades Weinberg’s Theorem by allowing the scalar field equation to be trivially solved at the Minkowski vacuum. This leads to a self-tuning evolution toward an attractor solution regardless of the value of the vacuum energy. However, the model relied on a carefully selected patchwork of potentials appearing in four Lagrangian terms (the ‘Fab Four’) to achieve a non-trivial cosmology \cite{copeland2012cosmology}. This work led to the re-discovery of the Horndeski theory \cite{horndeski1974second} as the most general scalar-tensor theory with a single scalar field interacting with the metric leading to at most second-order equations of motion making the entire class Ostrogradski stable. However, tuning to the Minkowski solution does not yield acceleration, and the mechanism did not allow for a matter-dominated era. Despite the limitations of the Fab-Four, it demonstrated that Horndeski theory provides a natural framework for exploring low-energy effective theories to understand mechanisms which may resolve the Cosmological Constant Problem.

More recently discovered `well-tempered' models \cite{appleby2018well}, \cite{emond2019well}, \cite{appleby2020well}, \cite{appleby2020wellfugue}, also based on Horndeski theories, tune the evolution to a late-time de Sitter attractor but leave the matter component untouched by requiring the Hubble evolution and scalar field equations to be proportional at the attractor. This appears to be a very fruitful approach to dynamical tuning which can lead to viable cosmological solutions with low effective energy-density despite the presence of a very high energy density vacuum energy. However, the well-tempered models \cite{appleby2018well}, \cite{emond2019well} found so far are restricted by the approach used to meet these criteria and seem to require complicated Lagrangians which obscure the underlying mechanisms. A reasonable question is to ask if the self-tuning and well-tempered mechanisms exhaust the possibilities for this approach, and if there exist simpler models which can result in behaviour consistent with observations?

In this paper, we present a new mechanism, differing from the self-tuning or well-tempered classes, to solve the Cosmological Constant Problem through a simple, minimal scalar field model. Our model belongs to the Kinetic Gravity Braiding (KGB) \cite{deffayet2010imperfect}, \cite{Pujolas_Sawicki_Vikman_2011}, \cite{Muharlyamov_Pankratyeva_2021} subclass of Horndeski theory. The mechanism contains two key mechanisms - cancellation of a large vacuum energy density through a linear potential, and dynamic self-tuning to a de Sitter attractor through a minimally modified kinetic energy and scalar field equation. We show the early-time cancellation of the vacuum energy during slow-roll allows for the existence of a matter dominated era after which the field enters stable, fast-roll evolution at the attractor. We also demonstrate the robustness of the mechanism under a phase transition in the vacuum energy density.

In Section \ref{sec:kgb}, we discuss the motivation for the KGB subclass and briefly review its properties. In Section \ref{sec:themodel}, we present our minimal model and its equations of motion. In Section \ref{sec:solutionstomodel}, we present our main numerical and analytical results. We also comment on the stability of the background spacetime. In Section \ref{sec:phasetrans}, we show the model copes with a phase transition in the vacuum energy density. In Section \ref{sec:energyscales}, we comment on the energy scales of our model, and in Section \ref{sec:summary}, we present our summary and concluding remarks.

\section{The Model Space: Kinetic Gravity Braiding}
\label{sec:kgb}

The most general, stable scalar-tensor model is the Horndeski theory \cite{horndeski1974second}, which we will assume is a suitable model space for a low-energy effective theory which solves the Cosmological Constant Problem. This model space is very large, but observational constraints substantially reduce the regime of viable theories. In particular, the stringent constraints from the speed of gravitational wave measurements \cite{lombriser2016breaking}, \cite{abbott2017gw170817}, \cite{baker2017strong}, \cite{creminelli2017dark}, \cite{ezquiaga2017dark} remove a large range of modified gravity models. This still allows for a modified gravitational coupling to the Ricci scalar in the Lagrangian, however even this term has come under significant pressure \cite{lombriser2016breaking}, \cite{lombriser2017challenges}, \cite{noller2019cosmological}. As a result, we shall assume a model space based only on Einstein gravity and a general k-essence and Galileon term which belongs to the Kinetic Gravity Braiding (KGB) subclass of Horndeski theory, which features an essential mixing or `braiding' of the scalar and metric terms in the equations of motion \cite{deffayet2010imperfect}. This class of models can exhibit many interesting properties such as attractor solutions, a Friedmann equation with self-interacting terms, and a scalar field that may respond to any form of external energy present. It has also been considered for the well-tempered models \cite{appleby2020well}, \cite{appleby2020wellfugue}.

The KGB gravitational action is given by
\begin{equation}\label{eq:action}
S = \int \smalld^4x \sqrt{-g}\left(\frac{\Mpl^2 R}{2} + \kessence(\phi,X) - G(\phi, X)\Box\phi \right) \, ,
\end{equation}
where $\kessence(\phi, X)$, a k-essence term, and $G(\phi, X)\Box\phi$, a generalised Galileon interaction term, are functions of the scalar field $\phi$ and its kinetic density $X= - g^{\mu\nu}\nabla_\mu\phi\nabla_\nu\phi/2$. Here, $\Mpl$ is the reduced Planck mass, $R$ is the Ricci scalar, and $\Box\phi = g^{\mu \nu} \nabla_{\mu} \nabla_{\nu} \phi$. The quantity in outer brackets is the Lagrangian density $\Lagr$ of the action, and we work in natural units where $[\Lagr] = M^4$.

The Friedmann equation and Hubble evolution equation can be derived from the energy-momentum tensor which has a diagonal form $T_\mu^{\;\nu}=(-\rho, p, p, p)$ on a flat, homogeneous and isotropic background given by $\smalld s^2 = -\smalld t^2 + a^2(t)\delta_{ij}\smalld x^i\smalld x^j$. For an action of the form Eq. \eqref{eq:action}, with a time-dependent scalar field $\phi = \phi(t)$, the Friedmann and Hubble evolution equations are given by \cite{kobayashi2010inflation}, \cite{bernardo2021self}
\begin{subequations}\label{eq:ogeq}
\begin{align}
3\Mpl^2 H^2 &= \rho \, ,
\\
-(3\Mpl^2 H^2 + 2\Mpl^2 \dot{H}) &= p \, ,
\end{align}
\end{subequations}
where $\rho$ and $p$ are the total energy density and total pressure. For the scalar field, the energy density and pressure are given by
\begin{subequations}\label{eq:densandpress}
\begin{align}
\rho_\phi &= 2\kessence_X X - \kessence + 3G_XH\phdt^3 - 2G_\phi X \, ,
\\
p_\phi &= \kessence - 2X(G_\phi + G_X \ddot{\phi}) \, .
\end{align}
\end{subequations}
The scalar field equation of motion can be evaluated from the expression $\nabla_\nu T_\mu^{\;\nu} = 0$, and can be found in \cite{kobayashi2010inflation}. In the above equations, we observe the importance of $G_X \neq 0$ which produces a term linear in $H$ in the Friedmann equation through the energy density, and a second-order time derivative of the field, $\ddot{\phi}$, in the pressure for a dynamical scalar field, which are essential to the self-tuning mechanism in our model.

\section{The Minimal Self-Tuning Model}\label{sec:themodel}

While the KGB theory defines our model space, we need to add additional constraints to find self-tuning models. Our guide here is to look for the simplest KGB model which has a cancelling potential and self-tunes towards an attractor. These criteria are satisfied by the k-essence term
\begin{subequations}\label{eq:kandg}
\begin{align}
\kessence(\phi, X) &= k(X) - V(\phi) \, ,
\\
k(X) &= - 3c_0 \frac{\hds}{M} X \, ,
\\
V(\phi) &= -c_1 M^3 \phi \, ,
\end{align}
\end{subequations}
where $k(\phi)$ is the  kinetic energy and $V(\phi)$ is a potential, and a Galileon term
\begin{equation}
G(X) = \frac{c_0}{M} \sqrt{2X}\, .
\end{equation}
The kinetic density is now $X=\dot \phi^2/2$, and in $G$ we choose the positive branch $\sqrt{2X}$ which simply reduces to $G \propto \phdt$ for a time-dependent scalar field $\phi(t)$. The dimensionless constants, $c_1$ and $c_0$, are chosen to explicitly separate the behaviour of the potential $V(\phi)$ from the rest of the field terms, respectively, in the equations of motion. $M$ is the mass scale of the scalar field, and $\hds$ is the attractor Hubble value. 

The $\kessence (\phi,X)$ term is split into a non-canonical kinetic energy term, $k(X),$ and a linear potential, $V(\phi)$. Despite the presence of a negative sign in $k(X)$, we shall find that the total kinetic energy is largely positive in our model as a result of `braiding' through the $G_X$ term. We will also show that the ghost-free condition is satisfied for the model in Section \ref{sec:solutionstomodel}. A linear potential of the form in Eq. \eqref{eq:kandg} can take on negative values for positive values of $\phi$, which will effectively cancel the large vacuum energy. It is also a crucial ingredient in well-tempered theories, with the `tempering' mechanism being difficult to achieve in shift-symmetric KGB theories without the linear potential \cite{bernardo2021self}. Here, we choose a linear potential out of simplicity, however in principle, any potential that can produce negative values should be able to cancel a positive $\rho_{\Lambda}$, but the dynamics of the field may differ. The $G(X)\Box\phi$ term is a form of generalised Galileon interaction. While the cubic Galileon model with $G \propto X\Box\phi$ has been ruled out through constraints from ISW, BAO and CMB data \cite{barreira2013nonlinear}, \cite{barreira2014observational}, \cite{renk2017galileon}, the particular form of $G$ in our model has not yet been compared with observations. A related model, mainly lacking the linear potential, was recently studied in the context of black holes in cubic Horndeski theories \cite{emond2020black}.

The action for our minimal model, with $\kessence$ and $G$ functions given by Eq. \eqref{eq:kandg}, is
\begin{equation}\label{eq:lagrangian}
S = \int \smalld^4x \sqrt{-g}\left(\frac{\Mpl^2R}{2} - 3c_0\frac{\hds}{M}X + c_1M^3\phi -\frac{c_0}{M} \sqrt{2X}\Box\phi + \rho_{\Lambda} + \Lagr_m \right) \, ,
\end{equation}
where we explicitly include the vacuum energy density $\rho_\Lambda$, and matter density through the matter Lagrangian $\Lagr_m$, along with our KGB terms. We note that the action possesses an effective symmetry under a shift, $\phi \to \phi + c$, where the constant offset from the linear potential can be absorbed by $\rho_{\Lambda}$. Some form of shift-symmetry may be useful in controlling quantum corrections \cite{padilla2015lectures}, \cite{appleby2018well} although it is unclear whether it is a crucial consideration in models of dark energy. A useful test would be to study the stability of the structure of the Lagrangian under matter loops.

The equations of motion for our minimal model, using Eqs. \eqref{eq:ogeq}, \eqref{eq:densandpress} and the scalar equation of motion, with  $\sqrt{2X}=\phdt$, are
\begin{equation}\label{eq:friedmann}
3\Mpl^2 H^2 = \rho_{m} + \rho_{\Lambda} + \frac{3c_0}{2M}(2H - \hds)\phdt^2 - c_1M^3\phi\, ,
\end{equation}
\begin{equation}\label{eq:hubble*}
-(3\Mpl^2 H^2 + 2\Mpl^2 \dot{H}) =  p_{m} + p_{\Lambda} - \frac{3c_0\hds}{2M}\phdt^2 - \frac{c_0}{M}\ddot{\phi} \phdt + c_1M^3\phi \, ,    
\end{equation}
\begin{equation}\label{eq:scalar}
(\ddot{\phi} + 3H\phdt)(H - \hds) + \dot{H}\phdt - \frac{c_1 M^4}{3c_0} = 0 \, .
\end{equation}
We can add Eq. \eqref{eq:friedmann} to Eq. \eqref{eq:hubble*} to give an evolution equation for the Hubble constant that  eliminates explicit $\Lambda$ dependence from the equation \cite{appleby2018well}. This is possible only because $p_{\Lambda} + \rho_{\Lambda} = 0$ for a constant vacuum energy density. Considering pressureless matter $p_m = 0$, we obtain the (modified) Hubble evolution equation
\begin{equation}\label{eq:hubble}
2\Mpl^2 \dot{H} = -\rho_{m} - \frac{3c_0}{M}\phdt^2(H - \hds) + \frac{c_0}{M}\phdt\ddot{\phi} \, .
\end{equation}
In addition, we have the continuity equation for pressureless matter in an evolving Universe 
\begin{equation}\label{eq:mattercontinuity}
\dot{\rho_m} + 3H\rho_m = 0 \, .
\end{equation}

We now infer some properties of the above equations and their implications. The background energy density of the scalar field expressed as the sum of its kinetic and potential parts, $\rho_\phi = K(\phdt,H) + V(\phi)$, can be deduced directly from the Friedmann Eq. \eqref{eq:friedmann} and is given by
\begin{equation}
\rho_\phi = \frac{3c_0}{2M}(2H - \hds)\phdt^2 - c_1 M^3\phi \, ,
\end{equation}
where we see that for $H\geq\hds$, the kinetic energy is positive and finite for a non-zero $\phdt$. The presence of a term linear in $H$ in the Friedmann equation allows for a positive kinetic energy despite the presence of a negative sign in $k(X)$. In Section \ref{subsec:stability}, we find that $H>\hds$ is a natural choice when considering the Null Energy Condition which requires that $\dot{H}\leq 0$. The potential can directly cancel the large vacuum energy density in the Friedmann equation; however it will do this dynamically. In the Hubble evolution Eq. \eqref{eq:hubble}, the presence of $\ddot{\phi}$ allows for a dynamic cancellation mechanism. The scalar field equation can be derived from the Friedmann, Hubble evolution and matter continuity equations and as such, is not independent. The scalar field Eq. \eqref{eq:scalar} can be solved exactly for $\phdt(t)$, to yield
\begin{equation}\label{eq:scalarsolution}
\phdt(t) = \frac{1}{(H - \hds)} \left( \frac{c_1M^4}{a^3c_0}\int a^3 dt \; + \; \frac{const}{a^3} \right) \, ,
\end{equation}
so that for a given form of scale factor $a(t)$, the behaviour of $\phdt(t)$ away from the attractor can be predicted. The first term in the brackets is a driving force due to the potential. It acts as a growing term during matter domination and approaches a constant value during exponential expansion. The constant of integration from the integral can be absorbed by the constant in the second term, which is a decaying term in an expanding universe.
The factor $1/(H-\hds)$ modifies the strength of the two terms as $H$ evolves toward the attractor $\hds$.
We note that the scalar field equation, and therefore its solution, is only valid away from the attractor. To deduce the behaviour of the field at the attractor we need to refer to other equations in the system.

The minimal model features a two-part mechanism - the stable cancellation of the large vacuum energy density, and self-tuning toward a de Sitter attractor. The former is primarily achieved by the linear potential which can directly offset a large $\rho_{\Lambda}$ by taking on negative values for positive $\phi$. The latter is largely a result of the `braided' kinetic energy density, with contributions from both the k-essence and Galileon terms, which can self-adjust to tune the Hubble constant to asymptotically arrive at the attractor solution. Even at the attractor a finite, positive kinetic energy remains, which allows the solution to be dynamically maintained by the interplay between the positive kinetic energy and negative potential energy.

We briefly contrast these equations to those of other self-tuning mechanisms, particularly that of the Fab-Four and well-tempered models. In Fab-Four, it is essential for self-tuning that the scalar field equation be trivially satisfied at the (Minkowski) attractor solution. For well-tempered tuning, the (modified) Hubble equation and scalar field equation need to be proportional at the attractor in the absence of matter. Clearly, neither of these conditions are satisfied for the above model, implying that the self-tuning mechanism for this system is different to those explored previously. In the next section, we elaborate on the model mechanism by presenting the numerical solutions to the equations of motion and considering their fast and slow-roll limits.

\section{Evolution of the Minimal Model}
\label{sec:solutionstomodel}

To study the evolution of the minimal model it is useful to re-write its equations of motion as dynamical equations for the Hubble parameter and the scalar field. It is also useful to move to dimensionless variables, using powers of the mass scale of the field $M$, for numerical purposes. We make the following substitutions (e.g., \cite{appleby2018well}) $h = H/M$, $\psi = \phi/M$, $\tau = M t$, $\mu = \Mpl^2/M^2$, $\rho_{m,0} = \rho_m/M^4$, and $\rho_{\Lambda,0} = \rho_{\Lambda}/M^4$. The attractor is now denoted by the dimensionless constant $\alpha = \hds/M$, and we represent $dx/d\tau =\, x'$.
The dimensionless evolution equations for $h'$ and $\psi''$, from Eqs. \eqref{eq:friedmann}, \eqref{eq:hubble}, and \eqref{eq:mattercontinuity}, are 
\begin{equation}\label{eq:hdot}
h' = \frac{- 3(h - \alpha)  \left(3c_0\psi'^2(2h - \alpha) + \rho_{m,0}\right) + c_1\psi'}{6\mu(h - \alpha) + 3c_0 \psi'^2} \, ,
\end{equation}
\begin{equation}\label{eq:psiddot}
\psi'' = \frac{9c_0\psi'(h-\alpha)(c_0\psi'^2 - 2\mu h) + 3c_0\psi'\rho_{m,0} + 2\mu c_1}{6\mu c_0(h - \alpha) + 3c_0^2\psi'^2} \, ,
\end{equation}
and the dimensionless matter continuity equation is given by
\begin{equation}\label{eq:continuitymatter}
\rho_{m,0}' + 3h\rho_{m,0} = 0 \, .
\end{equation}
The above three equations \eqref{eq:hdot} - \eqref{eq:continuitymatter} represent the full system that we can solve numerically through finite differencing. The equations are not dependent on $\rho_{\Lambda}$ explicitly, but $\rho_\Lambda$ influences the resulting dynamics through the Hamiltonian constraint during initialisation. Employing a similar technique to the original well-tempered paper \cite{appleby2018well}, we initialise $\psi$ through the Hamiltonian constraint Eq. \eqref{eq:friedmann} allowing us freedom to choose $(h_i, \psi'_i)$. We find that it is advantageous to be able to set the appropriate initial value for the Hubble constant $h_i = \sqrt{\rho_{m,0,i}/3\mu}$ for matter domination. We fix the following quantities for the results in this section, $M=10^{-3} \Mpl$, $\mu = 10^6$, $\alpha = 1$, $\rho_{m,0,i} = 10^9$, and $\rho_{\Lambda, 0} = 10^8$ which corresponds to a physical vacuum energy density value of $\sim 10^{69}\gev^4$. We also fix $c_0$ and $c_1$ to be equal to 1.
\par
\begin{figure}[h]
\centering
\includegraphics[width=0.75\textwidth]{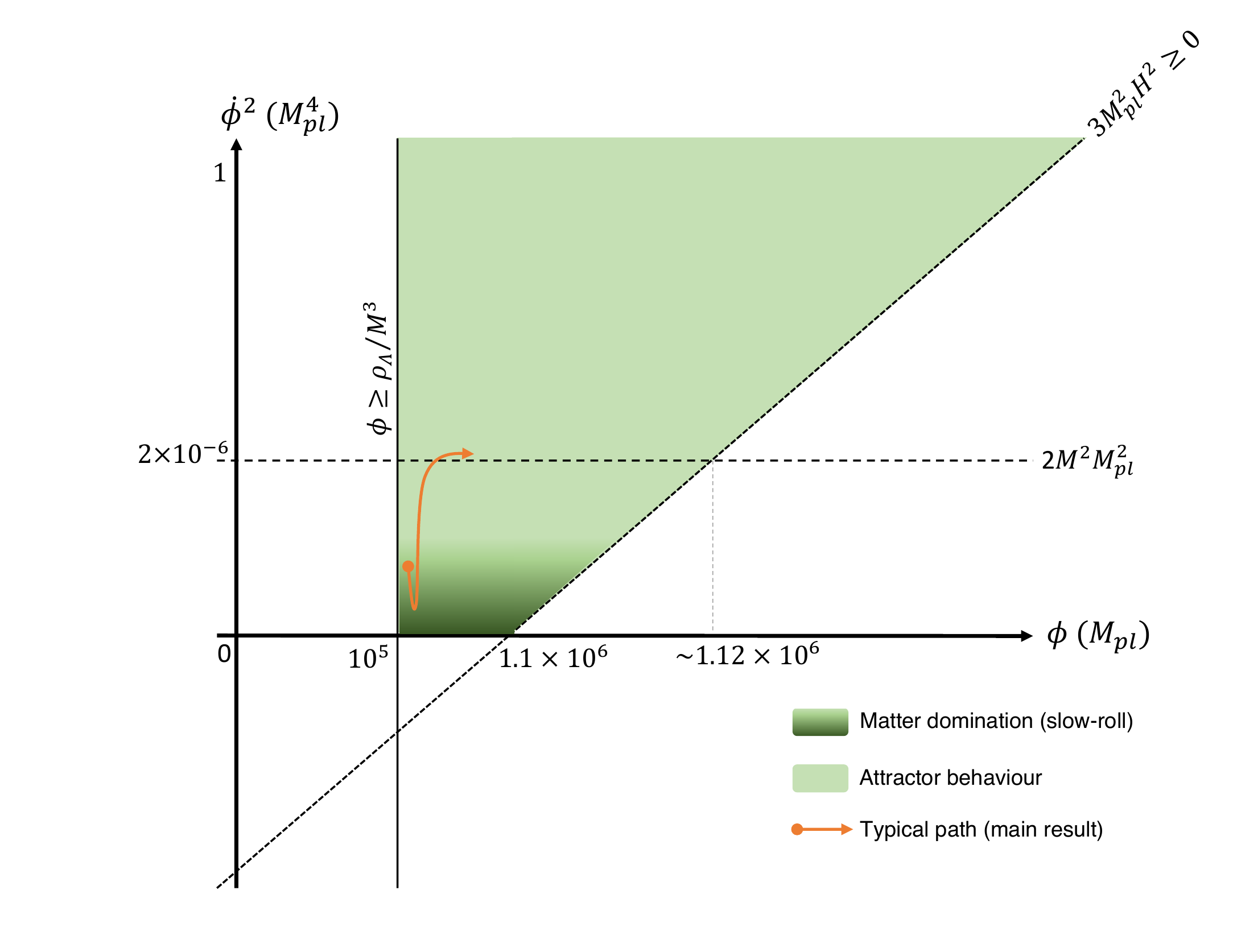}
\caption{In the above illustration the green region displays the initial values of $\dot{\phi}^2$ in ($\Mpl^4$) and $\phi$ (in $\Mpl$) that lead to attractor behaviour for $M=10^{-3}\Mpl$. The dark-green area illustrates the slow-roll region (marked by the dashed horizontal line) where conditions for matter-domination exist. The orange curve indicates the path for our main results.}
\label{fig:iniconditions}
\end{figure}

 It is also possible to initialise $h$ through the Hamiltonian constraint instead, giving one the freedom to choose the scalar field variables $(\psi'_i, \psi_i)$. Figure \ref{fig:iniconditions} illustrates the region of possible initial conditions in the dimensional field variable space $\phdt^2-\phi$ that lead to the de Sitter attractor at $\hds$, assuming $H_i > \hds$. This region is bound by the positivity of the square of field speed $\phdt^2$, the positivity of total energy $3\Mpl^2 H^2$, and the constraint that we require the initial value of the potential to be equal to or greater than the vacuum energy density, $|V_i|\geq \rho_\Lambda$. The plot illustrates that there is a wide range of values the potential may take to successfully cancel the vacuum energy density. The dark green region illustrates slow-roll values which allows for matter-domination in addition to an evolution to the attractor. Based on Figure \ref{fig:iniconditions}, this region is reasonably large, around $(\Delta\phdt)^{1/2}\times \Delta\phi \, \Mpl^2 = (2 \times 10^{-6} \Mpl^4)^{1/4} \times 10^6 \Mpl \approx 3 \times 10^4 \Mpl^2$.
\par
\begin{figure}[t]
\centering
\includegraphics[width=.75\textwidth]{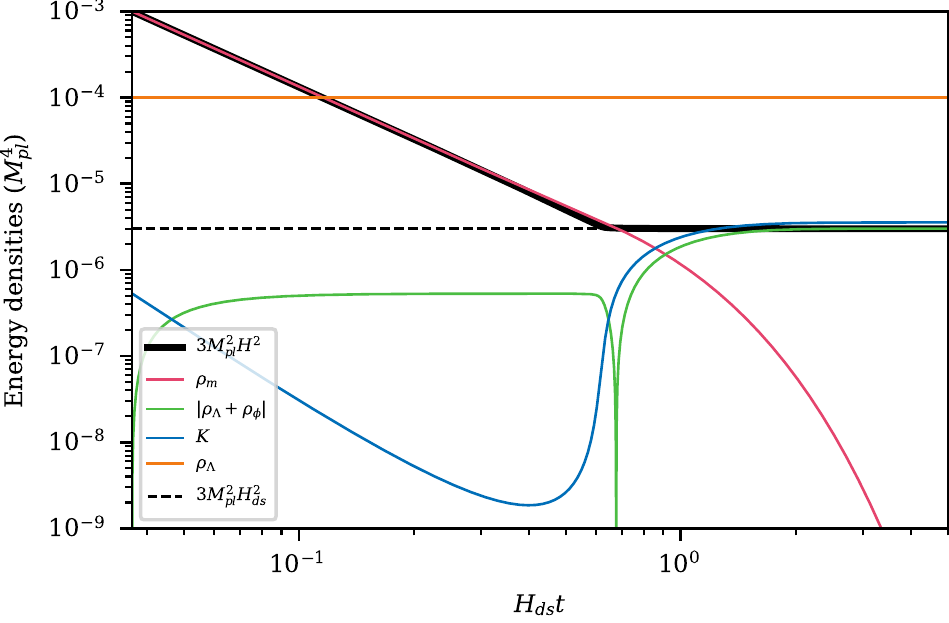}
\caption{\label{fig:energydensvstime} The evolution of energy densities, in units of $\Mpl^4$, for various components with time in dimensionless units of $\hds t$, for initially slow-roll values ($\psi_i' = 10^2$ or $\dot{\phi}_i = 10^{-4}\Mpl^2$.) $\rho_m$ is matter energy density, $\rho_\Lambda$ is vacuum energy density, $K$ is kinetic energy density of the field, $|\rho_\Lambda+\rho_\phi|$ is the absolute value of effective dark energy. The total energy density is $3M_{Pl}^2 H^2$ and the attractor energy density is given by $3M_{Pl}^2 H_{ds}^2$.}
\end{figure}

 Figure \ref{fig:energydensvstime} shows the numerical evolution of the various energy-densities in the model as a function of cosmic time scaled by the attractor, $H_{ ds} t$. The total energy density is given by $3M_{Pl}^2 H^2$ (solid black line). A large vacuum energy density $\rho_\Lambda$ (orange) is present, but it does not dominate the evolution of the model. The matter energy density, $\rho_m$ (pink), is seen to dominate at early times. We also show how the kinetic energy density of the scalar field, $K=3 c_0(2H-H_{ds})\dot\phi^2/(2M)$ (blue) and the effective dark energy density, $|\rho_\Lambda + \rho_\phi|$ (green), evolve. The dotted black line shows the attractor solution. We see that this model has a large vacuum energy, but the dynamic scalar field removes this, while allowing a matter dominated era which transitions into a scalar field dominated era with constant low-energy density. The final energy density is higher than observed, to allow numerical calculation. In Section \ref{sec:energyscales} we discuss suitable energy scales for the model which would reproduce observations. In the next section we discuss this behaviour in more detail using analytic solutions.
 
\subsection{Slow-roll Limit: Matter Dominated Era}

When in a slow-roll regime $(c_0 \psi'^2/2 \ll \mu)$, there is a decoupling of matter density $\rho_{m,0}$ with the Hubble attractor term $(h-\alpha)$ in Eq. \eqref{eq:hdot}. As a consequence, there is an absence of any screening of matter density by the attractor mechanism. As the kinetic energy is very small in this limit, the Hamiltonian constraint Eq. \eqref{eq:friedmann} allows the potential energy to cancel the vacuum energy density at initialisation. Under these conditions, matter is able to dominate the evolution of $h$, and Eq. \eqref{eq:hdot} simply reduces to 
\begin{equation}\label{eq:hgr}
h' \simeq -\rho_{m,0}/2\mu \, .
\end{equation}
The result is that at early times the Hubble parameter follows matter density evolution, generating a matter dominated era. 

\begin{figure}[h]
\centering
\includegraphics[width=.49\textwidth]{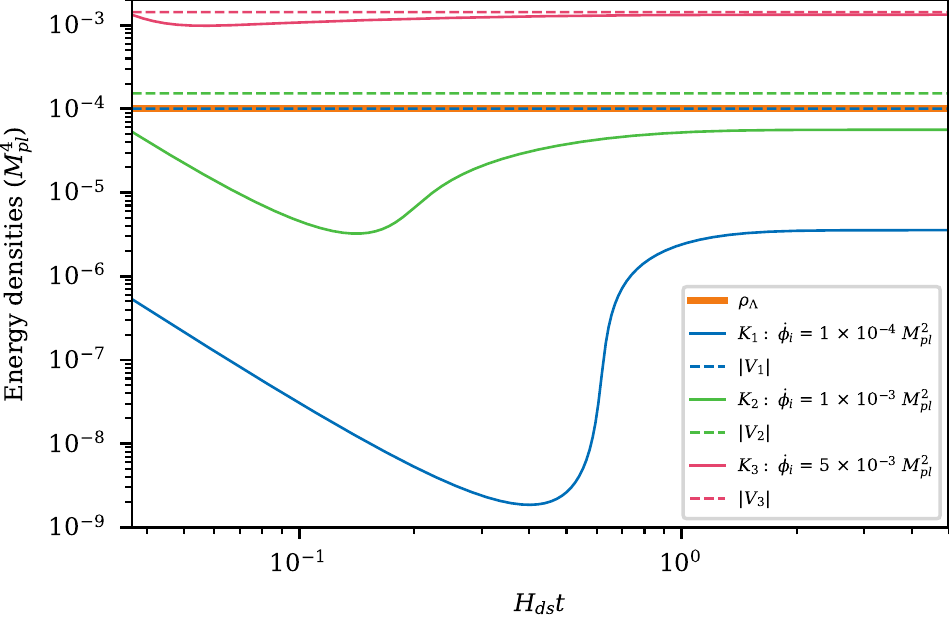}
\includegraphics[width =.49\textwidth]{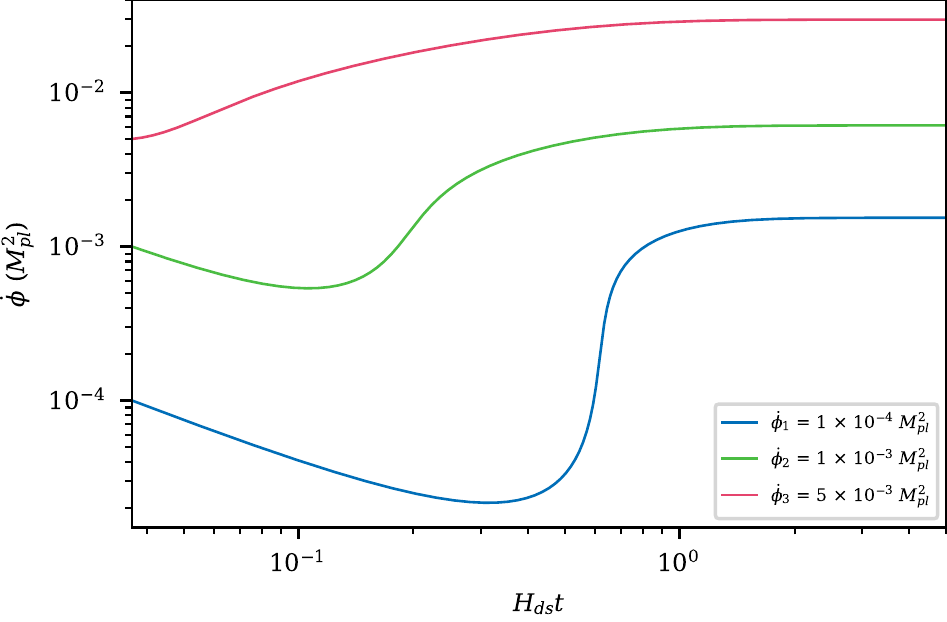}
\caption{\label{fig:kineticandpotential}LHS: This plot shows the evolution of kinetic energy density $K$ and the absolute value of potential energy density $|V|$ with time in dimensionless units (in $\hds t$) for different values of initial field speed $\phdt_i(=M^2\psi_i')$. Each colour, across the panels, corresponds to the same $\phdt_i$. RHS: This plot shows the corresponding evolution of the field speed $\phdt(=M^2 \psi')$ with time in dimensionless units (in $\hds t$). The initial value of Hubble constant, $H_i$, is fixed for all lines at $\sqrt{\rho_{m,i}/3\Mpl^2}$.}
\end{figure}

To understand the initial behaviour of the field, in particular its kinetic energy, $K$ (blue line in Figure \ref{fig:energydensvstime}), during matter domination, we can solve the scalar field equation \eqref{eq:scalarsolution} in its dimensionless form for $\psi'$ under the appropriate conditions of $a \propto \tau^{2/3}$ and $h = 2/3\tau$. The exact solution for $\psi'$ is given by
\begin{equation}\label{eq:psiprimeslowroll}
\psi' = \frac{A\tau^2 + B/\tau}{1 - 3\alpha\tau/2} \, ,
\end{equation}
where $A$ and $B$ are real constants. As discussed previously, the growing term is driven by the potential, while the decaying term is due the expansion, and the denominator modifies the effects of both as the attractor is approached.
At early times ($\tau \ll 1$), the decaying mode is dominant and $\psi'$ decreases with time as $1/\tau$. The scalar field kinetic energy density $K$, decays rapidly as $K \propto \tau^{-3} \propto a^{-9/2}$. In Figure \ref{fig:kineticandpotential} (right-panel), the gradual decay in $\dot\phi$ can be seen clearly for slow-roll (blue line). As $H$ is always decreasing, the combined decaying effect of $H$ and $\dot\phi$ can be observed in the kinetic energy in Figure \ref{fig:kineticandpotential} (left-panel) for slow-roll. 

As time progresses, the decay begins to slow down and the scalar field potential force term gains significance. The force term drives an increase in $\psi'$, the magnitude of which is driven even more rapidly by the shrinking denominator. The net result is that the effective dark energy density $|\rho_\Lambda + \rho_\phi|$, which was initially small and negative due to a small $K$, increases rapidly and crosses the zero barrier. Meanwhile, the total energy, $3\Mpl^2H^2$, continues to fall with $\rho_m$ until it reaches close to the attractor value at $3 \Mpl^2 \hds^2$. Once matter dilutes away further and the total energy approaches the attractor value, the scalar field equation \eqref{eq:scalar} no longer holds, and the evolution of the Hubble constant transitions to fast-roll. The fast-roll solution given by Eq. \eqref{eq:largepsiprimeexact} is detailed in the next subsection. To summarise its effect, in the fast-roll limit, $h$ asymptotes to the de Sitter attractor at $\alpha(= \hds/M)$, which results in an accelerated expansion at an energy scale set by the attractor energy value $3\Mpl^2 \hds^2$, much lower than the scale of vacuum energy density. The fast-roll solution for $\psi''$ asymptotically approaches $0$ in the attractor limit, explaining the flatness of $\psi'$ at late times for all field speed and kinetic energy lines in Figure \ref{fig:kineticandpotential}. The attractor solution is maintained by the kinetic energy of the field. As the field gradually rolls down the negative potential slope, the total energy receives small, negative contributions from the potential which are balanced by the self-adjusting behaviour of the kinetic energy. 

\subsection{Fast-roll Limit: Imperfect Cancellation}
\begin{figure}[h]
\centering
\includegraphics[width=.75\textwidth]{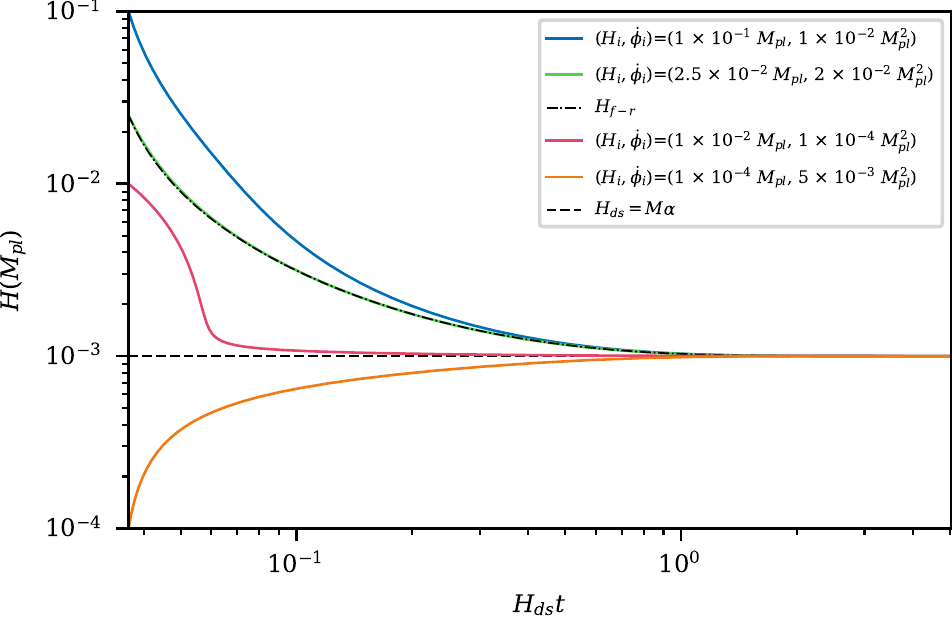}
\caption{\label{fig:hubblevstime} The evolution of the Hubble constant $H (= h/M)$ (in $\Mpl$) with time in dimensionless units (in $\hds t$) towards the attractor at $\hds(=M\alpha)$ for different initial values of $(H_i, \phdt_i)$. The analytical solution for the Hubble constant in fast-roll $H_{f-r}$, given by Eq. \eqref{eq:largepsiprimeexact}, is also plotted along with its corresponding numerical solution.}
\end{figure}

In fast-roll ($c_0\psi'^2/2 \gg \mu$), $\psi'$ dominates the evolution equations \eqref{eq:hdot}-\eqref{eq:psiddot}, and they simplify to
\begin{subequations}\label{eq:largepsiprimelimit}
\begin{align}
\label{eq:largepsiprimelimit:h}
h' &= - 3(h-\alpha)(2h - \alpha) \, ,
\\
\label{eq:largepsiprimelimit:psiprime}
\psi'' &= 3\psi'(h-\alpha) \, ,
\end{align}
\end{subequations}
and can be exactly solved for $h$ and $\psi'$ to give their fast-roll solutions
\begin{subequations}\label{eq:largepsiprimeexact}
\begin{align}
h_{f-r} &= \alpha \left( \frac{e^{3\alpha \tau} - c}{e^{3\alpha \tau} - 2c} \right) \, ,
\\
\psi_{f-r}' &= \psi'_0 \sqrt{1 - 2c\exp(-3\alpha \tau)} \, .
\end{align}
\end{subequations}
Here, $c$ and $\psi'_0$ are both positive constants that can be determined from initial conditions. The fast-roll analytical solution for the Hubble value $H_{f-r}(=h_{f-r}M)$ (black dash-dot line) is plotted with its full numerical solution (green) in Figure \ref{fig:hubblevstime}. The above solution for $h_{f-r}$ evolves to $\alpha$, asymptotically, at late times. In this limit, we also find that $\psi'' \to 0$ as $h\to\alpha$. 

As discussed in the previous subsection, slow-roll can transition to a fast-roll solution near the attractor. Therefore, we find that for both limits, the Hubble attractor solution is asymptotically approached and that $\psi'$ flattens at the attractor. We also plot the evolution of $H$ for different initial values in Figure \ref{fig:hubblevstime} to validate the attractor solution.

We note that though we call this region fast-roll, $\phdt$ is small compared to the Planck scale. This ensures that Planck-suppressed EFT operators such as $R_{\mu\nu}\nabla^\mu\phi\nabla^\nu\phi/\Mpl^2$ remain sensible.

We now elaborate on the energy picture in this limit. With fast-roll initial conditions, the initial value of the potential energy is set very large and negative through the Hamiltonian constraint Eq. \eqref{eq:friedmann}, such that $|V_i| \gg \rho_\Lambda$. The significant kinetic energy of the field $K$ drives an increase in $\psi'$. At the same time, $h$ decreases during its evolution. In Figure \ref{fig:kineticandpotential} (left-panel), we see that at early times, the net effect on $K$ (pink solid line) is that it initially decreases, causing the large potential (dashed pink line) to become dominant and over-cancel the matter energy density, giving an overall suppressed evolution of $3\Mpl^2 H^2$ compared to that of matter density. Therefore, due to the imperfect cancellation of vacuum energy density that tunes away some of the matter density, it is not possible to have a matter dominated era in fast-roll. Despite the undesirable tuning of $\rho_m$, fast-roll is still successful in self-tuning to the de Sitter attractor.
\subsection{Stability}
\label{subsec:stability}
To assess the stability of the background spacetime in this model, we evaluate and briefly discuss the Null Energy Condition, Laplace and ghost-free conditions, and the sound speed squared of scalar perturbations. It is generally possible for KGB models to violate the Null Energy Condition (NEC). For the energy-momentum tensor, $T_{\mu\nu}$, the NEC is given by $T_{\mu\nu}n^{\mu}n^\nu \geq 0$ for any null vector $n^\mu$. For a flat-FLRW spacetime the condition simply reduces to $\rho + p \geq 0$, where $\rho$ and $p$ are the total energy density and pressure of the Universe \cite{rubakov2014null}. Consequentially, this implies that for the NEC to be satisfied, $\dot H \leq 0$ always. In our model, we find that the condition is naturally satisfied when $H_i > H_{ds}$ due to the attractor behaviour (see Figure \ref{fig:hubblevstime}). It is interesting to note that NEC violation may not necessarily make a model non-viable, and it is possible for a scalar field model to stably violate the NEC by ensuring that the background spacetime is free of pathologies \cite{rubakov2014null}, \cite{dubovsky2006null}.

The ghost-free $\ghostfree > 0$ and Laplace $\laplace > 0$ conditions for a KGB model with $\kessence(\phi, X)$ and $G(X)$ are of the form \cite{appleby2018well}, \cite{kobayashi2010inflation}
\begin{equation}
\laplace = \kessence_X + 2G_X(\ddot{\phi} + 2H\phdt) - 2X^2 \frac{G^2_X}{\Mpl^2} + 2G_{XX}X\ddot{\phi} \, ,
\end{equation}
\begin{equation}
\ghostfree = \kessence_X + 2X\kessence_{XX} + 6G_XH\phdt + 6X^2\frac{G^2_X}{\Mpl^2} + 6G_{XX}HX\phdt \, ,
\end{equation}
and can be written for our model in dimensionless units as
\begin{equation}\label{eq:modellaplace}
\laplace = c_0 (4h - 3\alpha) - \frac{c_0^2 \psi'^2}{2\mu} + \frac{c_0 \psi''}{\psi'} \, ,
\end{equation}
\begin{equation}
\ghostfree = 3c_0 (h - \alpha) + \frac{3c_0^2 \psi'^2}{2\mu} \, .
\end{equation}
We first note that for $h\geq\alpha$, $\ghostfree$ is always positive for our model and is therefore satisfied for our main results. Our background is ghost-free despite the presence of a negative sign in the kinetic term $k(X)$ in the Lagrangian. 

The behaviour of $\laplace$ is less straightforward and so, we consider the slow-roll limit of $\laplace$ at the attractor for simplicity. Near the attractor, $\psi' \sim 10^3$, making the term $c_0 \psi''/\psi'$ very small. The Laplace condition simplifies to $\laplace \to c_0\alpha - c_0^2 \psi'^2/2\mu$ which for the above values is $\approx 1/2$. So, $\laplace > 0$ is so far satisfied for the model near the attractor. We now note that in the future, we expect $\psi'$ to increase due to a small but finite and positive $\psi''$ and eventually, $\laplace$ may become negative in the future. This may also have implications for the sound speed squared of scalar perturbations, which is simply given by $\sss = \laplace/\ghostfree$ \cite{kobayashi2010inflation}, which needs to be positive for the controlled growth of perturbations. Ideally, we would also require sub-luminal propagation for causality. To that effect, we need $0 \leq \sss \leq 1$. In the attractor limit as above, this requires that at the attractor the field speed obeys $2 \mu\alpha/c_0 \geq \psi'^2 \geq \frac{1}{2} \mu\alpha/c_0$. This is satisfied at the attractor; however the field speed will gradually continue to increase and $\sss$ will become negative as a result. We note that these instabilities only arise several $\hds t$ after reaching the attractor. This indicates the presence of possible future pathologies in the background de Sitter spacetime and requires further exploration. 

\section{A Phase Transition in Lambda}\label{sec:phasetrans}
\begin{figure}[h]
\centering
\includegraphics[width = 0.75\textwidth]{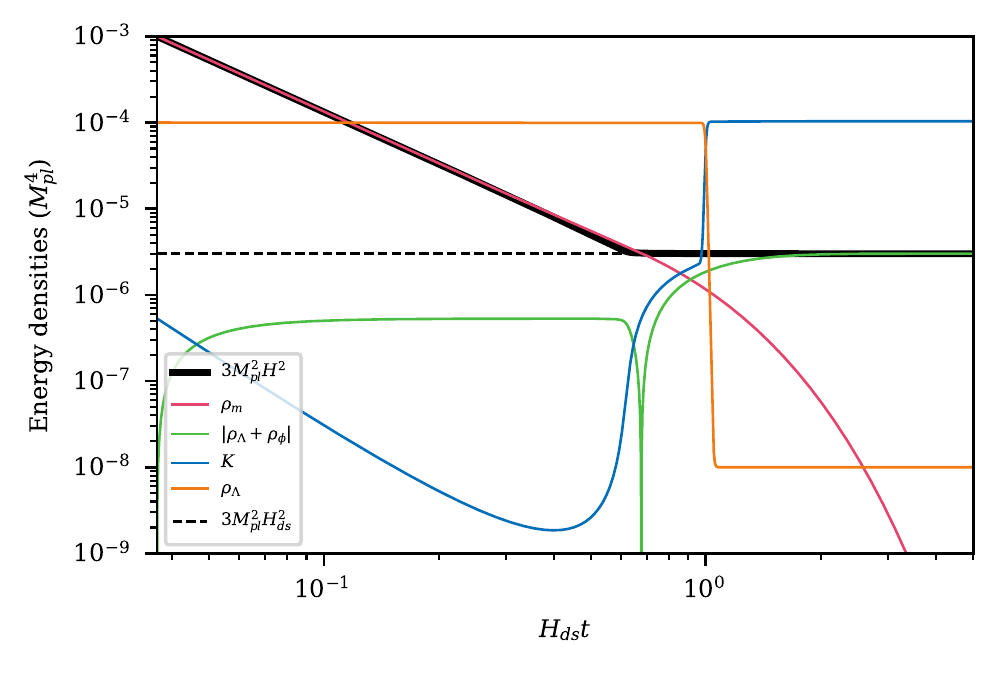}
\caption{A phase transition modelled by Eq. \eqref{eq:phasetrlambda} in vacuum energy density $\rho_\Lambda$ with $\rho_1=10^{-4}\Mpl^4$ and $\rho_2 =10^{-8}\Mpl^4$, at $ H_{ds} t = \tau_c = 1.0$. The total energy is given by $3M_{Pl}^2 H^2$, the matter energy density is $\rho_m$, the absolute value of effective dark energy density is given by $|\rho_\Lambda + \rho_\phi|$, the kinetic energy density of the field is $K$, and the attractor energy value is given by $3 M_{Pl}^2 H_{ds}^2$. Here, $\rho_{m,i} = 10^{-3} \Mpl^4$, $H_i = \sqrt{\rho_{m,i}/3\Mpl^2}$, $\phdt_i = 10^{-4} \Mpl^2$ for slow-roll and $\tau_0 =0.01$.}
\label{fig:changinglambda}
\end{figure}

The vacuum energy density may undergo phase transitions during different epochs in the Universe's history. In this section, we assess the stability of the attractor mechanism in response to a rapid change in the vacuum energy density. We model the phase transition in $\rho_{\Lambda}$ with the following switch equation
\begin{equation}\label{eq:phasetrlambda}
\rho_{\Lambda, 0} = \left( \frac{\rho_1 + \rho_2}{2}\right)  - \left( \frac{\rho_1 - \rho_2}{2}\right) \tanh \left(\frac{\tau - \tau_c}{\tau_0} \right) \, .
\end{equation}
The parameters $\rho_1$ and $\rho_2$ define the initial and final phase transition values of vacuum energy density respectively. $\tau_c$ is the time at which the transition takes place, and $\tau_0$ describes the period over which the transition occurs. Since a change in the vacuum energy is reflected in a change in its pressure, the evolution of vacuum energy density obeys a continuity equation similar to that for the matter component and is given by
\begin{equation}\label{eq:lambdacontinuity}
\rho_{\Lambda, 0}' + 3h(\rho_{\Lambda, 0} + p_{\Lambda, 0}) = 0 \, .
\end{equation}
During the transition, the vacuum energy density is no longer constant and $p_{\Lambda,0} \neq - \rho_{\Lambda,0}$. As the (modified) Hubble equation \eqref{eq:hubble} no longer holds, we solve the system of equations formed by Eqs. \eqref{eq:friedmann}, \eqref{eq:scalar}, \eqref{eq:mattercontinuity}, and \eqref{eq:lambdacontinuity}. The phase transition and the response of the scalar field is illustrated in Figure \ref{fig:changinglambda} for slow-roll $(c_0 \psi'^2/2 \ll \mu)$.

Before the phase transition occurs, the numerical behaviour of the system is identical to that of slow-roll. To summarise this briefly, slow-roll conditions allow cancellation of the vacuum energy density through the tadpole potential and generate a matter dominated era. The kinetic energy of the field $K$ is very small during matter domination and therefore, $\phi$ is slow moving. Near the attractor, the field enters a fast-roll solution to drive the total energy $3M_{Pl}^2 H^2$ to its attractor value $3 M_{Pl}^2 H_{ds}^2$. Without a phase transition in the vacuum energy density $\rho_\Lambda$, the kinetic energy would flatten at this stage to asymptote to a constant value. However, in the presence of a phase transition the value of $\rho_{\Lambda}$ falls from $10^{-4}\Mpl^4 \to 10^{-8}\Mpl^4$. In the absence of a self-tuning mechanism, the potential energy density would be too large and negative in the absence of an equally large vacuum energy density and move the Hubble value off the attractor. However, we find that the scalar field responds to the phase transition by quickly increasing its kinetic energy $K$ by several orders of magnitude in order to balance the large potential energy. The self-tuning mechanism is therefore able to maintain the attractor solution, making the mechanism stable under a phase transition in $\rho_\Lambda$. We find that the mechanism is successful even if the phase transition occurs before $H$ has reached the attractor solution.

\section{Energy Scales}\label{sec:energyscales}
We have shown that it possible to find a simple, self-tuning scalar field cosmological model which can remove a high-energy quantum vacuum energy component and re-tune to a low-energy attractor solution. However, we have so far illustrated this mechanism assuming sub-Planckian vacuum energy density $10^{-4} \Mpl^4$ and reduced this by a factor of $\sim 30$ to allow numerically viable solutions. However, our model can be scaled to any vacuum energy density. To match observations, the attractor energy $3 M_{Pl}^2 H_{ds}^2(= 3 \mu \alpha M^4)$ should coincide with the scale of the critical density today $10^{-47} \gev^4$. Considering that we fix $\alpha=1$ for the model to keep the coefficient of the kinetic operator $\mathcal{O}(1)$, we find the mass of the scalar field would need to be $M \sim 10^{-33} \, \text{eV} \sim 10^{-60} \Mpl$. The mass corresponds to the current Hubble scale $H_0 \sim 10^{-33}\, \text{eV}$, and is similar to that proposed for quintessence \cite{tsujikawa2013quintessence} and ultralight scalar field dark matter models, reflecting a low-energy particle physics scale.

The scaling of our model is also non-trivially dependent on the amplitude of the vacuum energy. Detailed QFT calculations suggest that the renormalised value of the vacuum energy density may be much smaller than the value simply calculated using the Planck mass as cut-off \cite{martin2012everything}, as much as 60 orders of magnitude smaller. A smaller value of vacuum energy density will directly affect the choice of scaling of parameters in the model. An implication of the small mass scale for scalar field models where the potential $V\propto M^3\phi$ removes a large vacuum energy density with a very small $M$ is that the scalar field value $\phi$ is required to be very large. For a vacuum energy density of $10^{-4} \Mpl^4$ with a mass scale $10^{-60}\Mpl$, the typical field value required would be $\phi \sim 10^{176} \Mpl$. Such large field values can correspond to a breakdown of low energy effective field theory \cite{klaewer2017super}, but this may be protected by shift symmetry. Exact global symmetries are forbidden according to quantum gravity conjectures \cite{banks2011symmetries}, but approximate symmetries may alleviate these tensions.

\section{Summary and Conclusions}\label{sec:summary}
The Cosmological Constant Problem, which tries to reconcile the large vacuum energy required by quantum field theory and the observed small, but non-zero, dark energy needed to accelerate the expansion of the Universe, has proven stubborn. In particular, standard renormalisation techniques do not resolve the discrepancy and Weinberg’s No-Go Theorem blocks mechanisms to remove the vacuum energy with a scalar field relaxing to a static state in Minkowski spacetime.

In this paper we have developed a new, minimal scalar field model which avoids Weinberg's No-Go Theorem by using a dynamic field which can both cancel the large vacuum energy and self-tune to a de Sitter attractor. We have shown that the model is able to preserve a matter era during slow-roll and produce accelerated expansion at a scale much lower than one credited to the vacuum energy density alone. We have also shown that the attractor solution is valid, and stable under a sudden phase transition in the value of vacuum energy density. We have, therefore, shown that there exists a minimal, self-tuning model that can reproduce the observable Universe with just a few ingredients.

Our model builds upon previous work to find self-tuning or well-tempered behaviour, but does not belong to either type of model and appears to be the simplest example of this type of mechanism. Under the constraints we have imposed, the model we find belongs to the Kinetic Gravity Braiding sub-class of Horndeski scalar-tensor theory and has not been excluded by observations. As well as a viable mechanism to alleviate the Cosmological Constant Problem, we have shown there exists a wider class of models which can produce this mechanism. The model seems to require a very light scalar field mass, but uncertainty in the theoretical value for vacuum energy may increase this value. The attractor energy in our model is not itself predicted by the model but defined to match the observed value of dark energy and therefore, reflects low-energy physics scales. We conclude that this mechanism provides a promising route to a complete solution of the Cosmological Constant Problem.
\appendix

\acknowledgments
A.K. is grateful to the University of Edinburgh for the PCDS and EGRS PhD scholarships. A.T thanks STFC for a Consolidated Grant.

\bibliography{references} 
\bibliographystyle{ieeetr}
\end{document}